\documentstyle[preprint,aps]{revtex}
\newcommand{\be}{\begin{equation}} \newcommand{\ee}{\end{equation}}
\newcommand{\bea}{\begin{eqnarray}}\newcommand{\eea}{\end{eqnarray}}

\begin{document}
\draft
\preprint{IP/BBSR/95-67}
\title { Topological and Nontopological Solitons in a Gauged O(3) Sigma
Model with Chern-Simons term}
\author{Pijush K. Ghosh$^{*}$ and Sanjay K. Ghosh$^{**}$}
\address{Institute of Physics, Bhubaneswar-751005, INDIA.}
\footnotetext {$\mbox{}^*$ E-mail:
pijush@iopb.ernet.in \\ $\mbox{}^{**}$ E-mail: sanjay@iopb.ernet.in}
\maketitle
\begin{abstract}
The $O(3)$ nonlinear sigma model with its $U(1)$ subgroup gauged,
where the gauge field dynamics is solely governed by a Chern-Simons term,
admits both topological as well as nontopological self-dual soliton
solutions for a specific choice of the potential.
It turns out that the topological solitons are infinitely
degenerate in any given sector.
\end{abstract}
\narrowtext

\newpage

The $O(3)$ sigma model in $2+1$ dimensions is exactly integrable \cite{sig}
in the Bogomol'nyi limit \cite{bogo}. The stability of these soliton
solutions are guarantied
by topological arguments. However, the solitons in this model, which can be
expressed in terms of rational functions, are scale invariant. Due to this
conformal invariance, the size of these solitons can change arbitrarily
during the time evolution without costing any energy. In fact, numerical
simulation of these soliton solutions indeed supports such a behaviour
\cite{ns}. Naturally, the particle interpretation of these solitons upon
quantization is not valid. There are several ways to break the scale
invariance of this model \cite{sky,qlump}. Construction of
Q-lumps \cite{qlump} is one such example
where the scale invariance is broken by including a specific
potential term in the sigma model. The collapse of the soliton's size
in this model is prevented by making a rotation in the internal space of
the field variables. These finite energy solitons are necessarily
time-dependent with a constant angular velocity. Very recently, it was
shown that the scale invariance of the $O(3)$ sigma model can also be
broken by gauging the $U(1)$ subgroup as well as including a
potential term \cite{dur}.
However, in contrast to the Q-lump case, no rotation in the internal
space of the scalar field variables is necessary. These soliton solutions
are static with zero charge and angular momentum and though the energy is
quantized, flux is not. It is worth enquiring at this point whether or not
static soliton solutions with nonzero but finite charge and angular momentum
is possible in any version of gauged $O(3)$ sigma model.

In this context it is worth recalling that static solitons in $2+1$
dimensional abelian Higgs model acquire
nonzero charge and angular momentum in the
presence of the Chern-Simons ( CS ) term \cite{paul}.
The purpose of this letter is to show that the gauged $O(3)$ sigma model
with the gauge field dynamics governed solely by a CS term indeed admits
soliton solutions with broken scale invariance.
To put it in another way,
in gauged sigma model with pure CS term one can study the breaking of
scale invariance of the solutions due to two simultaneous remedies,
(i) gauging of the $U(1)$ subgroup and (ii)
making the static solitons spin in the internal space. The study of soliton
solutions in gauged
sigma model with a CS term is also well motivated due to its possible
relevance in planar condensed matter systems where a charge-flux
composite obeying fractional statistics plays a major role \cite{poly}.

We show that the specific
form of the scalar potential which is required in order to have
Bogomol'nyi bound \cite{bogo}
allows us to have two different kinds of soliton solutions. In one case the
stability is guarantied by topological arguments, while for the other
no topological criteria can be made to establish the stability of the
solitons. The behaviour of the field variables for the later one is very
similar to the self-dual nontopological vortices in pure CS
theory \cite{hong,jlw}. Hence, we refer them throughout this paper as
nontopological solitons.  As far as we are aware off, this is the first
instance when both topological and nontopological soliton solutions
simultaneously exist in modified $O(3)$ sigma model. The
flux, charge and angular momentum is not quantized for either the topological
or the nontoplogical solitons. However, the energy is quantized in case of
topological solitons, while it is not quantized for the nontopological
solitons.
In particular, soliton solutions in any given topological sector with
degree $N$
have same energy but different charge, flux and angular momentum
characterized by a parameter $\beta_1$ ( to be defined below ) which
continuously interpolates within the range $0 < \beta_1 < 1 -
\frac{1}{2 N}$. Consequently, the
topological solitons are infinitely degenerate in each sector.
Both the energy density and the magnetic field in
this model for nontopological solitons are concentrated around two
concentric rings with different radii.
The magnetic field has doubly degenerate maxima, while
the energy density has two non-degenerate maxima.

Let us consider the following Lagrangian,
\be
{\cal{L}} = \frac{1}{2} D_\mu {\vec \phi} . D^\mu {\vec \phi} +
 \frac{\kappa}{4}
\epsilon^{\mu \nu \lambda} A_\mu F_{\nu \lambda}
-\frac{1}{2 \kappa^2} \left ( 1 + {\hat{n}}_3 . {\vec \phi} \right ) \left ( 1
- {\hat{n}}_3 . {\vec \phi} \right )^3
\label{eq0}
\ee
\noindent where ${\vec \phi}$ is a three component vector ${\vec \phi}=
\phi_1 {\hat{n}}_1 + \phi_2 {\hat{n}}_2 + \phi_3 {\hat{n}}_3 $ with
unit norm, i.e. $ {\vec \phi} . {\vec \phi} = 1 $, in the internal
space spanned by the three unit vectors ${\hat{n}}_1$, ${\hat{n}}_2$ and
${\hat{n}}_3$. We work here in Minkowskian space-time with the signature
$g_{\mu \nu}=( 1, -1, -1 )$. The velocity of light $c$ and the
Planck's constant in units of $\frac{1}{2 \pi}$ are taken to be unity.
In this case the coefficient of the CS term ($\kappa$) has dimension of
the inverse mass. The factor $\frac{1}{2 \kappa^2}$ in front of the
potential term in (\ref{eq0}) is chosen so as to have a Bogomol'nyi bound. The
soliton solutions in (\ref{eq0}) can be studied away from the Bogomol'nyi
limit by replacing this factor with any constant having mass dimension $2$.

The covariant derivative $D_\mu {\vec \phi}$ is defined as
\be
D_\mu {\vec \phi} = \partial_\mu {\vec \phi} +
A_\mu {\hat{n}}_3 \times {\vec \phi} .
\label{eq1}
\ee
\noindent The Lagrangian ( \ref{eq0} ) is invariant under a $SO( 2 )$
iso-rotation around the axis ${\hat{n}}_3$. In fact, one can use the identity
$D_\mu {\vec \phi} . D^\mu {\vec \phi} = {\mid (\partial_\mu + i A_\mu)(\phi_1
+ i \phi_2)\mid}^2 + \partial_\mu \phi_3 \partial^\mu \phi_3$ to see the local
$U(1)$ nature of it. The potential has two degenerate minima at $\phi_3
=\pm 1$. The constraint  $ {\vec {\phi}} . {\vec {\phi}} = 1 $
essentially implies that $\phi_1$ and $\phi_2$ are zero in case
$\phi_3 = \pm 1$. As a result, the local $SO(2)$ ( or $U(1)$ ) symmetry
is not broken
spontaneously. Note that the gauge field dynamics is solely governed
by a CS term. This is justifiable in the long wave length limit where
the Maxwell term being a double derivative term ( compared to the CS
term ) can be dropped from the action.

The equations of motion which follow from (\ref{eq0}) are
\be
D_\mu {\vec J}^\mu = - \frac{1}{\kappa^2} ({\hat{n}}_3 \times {\vec \phi})
(1 - {\hat{n}}_3 . {\vec \phi})^2 (1 + 2 {\hat{n}}_3 . {\vec \phi})
\label{eq2}
\ee
\be
j^\mu = \frac{\kappa}{2} \epsilon^{\mu \nu \lambda} F_{\nu \lambda}
\label{eq3}
\ee
\noindent where the current ${\vec J}^\mu$ is defined as
\be
{\vec J}^\mu = {\vec \phi} \times D^\mu {\vec \phi}
\label{eq4}
\ee
\noindent and the $U(1)$ current is $j^\mu = - {\vec J}^\mu . {\hat{n}}_3$.
In obtaining
Eq. (\ref{eq2}) the constraint  $ {\vec {\phi}} . {\vec {\phi}} = 1 $ is
taken care of by use of a Lagrange multiplier though not mentioned
explicitly in (\ref{eq0}). The zero component of Eq. (\ref{eq3}), i.e.
the Gauss law implies that the field configurations with nonzero magnetic
flux $\Phi$ essentially carry nonzero $U(1)$ charge $Q= - \kappa \Phi$.

The energy functional $E$ can be obtained by varying (\ref{eq0}) with
respect to the background metric and CS term being a topological term
does not contribute to it,
\be
E = \frac{1}{2} \int d^2 x \left [ \left ( D_1 {\vec \phi} \right )^2
+ \left ( D_2 {\vec \phi} \right )^2 +
\frac{\kappa^2 F_{12}^2}{\phi_1^2+\phi_2^2}+
\frac{1}{ \kappa^2} ( 1 + {\hat{n}}_3 . {\vec \phi} ) ( 1 -
{\hat{n}}_3 . {\vec \phi} )^3 \right ] .
\label{eq5}
\ee
\noindent The potential $A_0$ has been eliminated by using the Gauss
law. The energy functional (\ref{eq5}) can be rearranged as
\bea
E \ & = & \ \frac{1}{2} \int d^2x \left [ \left ( D_i {\vec \phi}
\pm \epsilon_{ij} {\vec \phi} \times D_j {\vec \phi} \right )^2 +
\frac{\kappa^2}{1 -
\phi_3^2} \left ( F_{12} \mp \frac{1}{\kappa^2} \left ( 1 + \phi_3 \right )
\left ( 1 - \phi_3 \right )^2 \right )^2 \right ]\nonumber \\
& & \ \pm 4 \pi \int d^2x K_0 , \ \ \ i, j = 1, 2
\label{eq6}
\eea
\noindent where $K_0$ is the zero component of the topological current $K_\mu$
defined as,
\be
K_\mu = \frac{1}{8 \pi} \epsilon_{\mu \nu \rho} \left [ {\vec \phi} .
D^\nu {\vec \phi} \times D^\rho {\vec \phi} + F^{\nu \rho} \left ( 1 -
{\hat{n}}_3 . {\vec \phi} \right ) \right ] .
\label{eq7}
\ee
\noindent The energy in Eq. (\ref{eq6}) has a lower bound
$E \geq 4 \pi T$ in terms of the topological charge $T= \int d^2x K_0$.
The bound is saturated when the following Bogomol'nyi equations are
satisfied,
\be
D_i {\vec \phi} \pm \epsilon_{ij} {\vec \phi} \times D_j {\vec \phi} = 0 ,
\label{eq8}
\ee
\be
F_{12} \mp \frac{1}{\kappa^2} \left ( 1 + \phi_3 \right )
\left ( 1 - \phi_3 \right )^2 = 0 .
\label{eq9}
\ee
\noindent One can
check that these Bogomol'nyi equations are consistent with the second
order field equations (\ref{eq2}).

Using the stereographic projections,
\be
u_1 = \frac{\phi_1}{1+\phi_3}, \\\\\\\ u_2 = \frac{\phi_2}{1+\phi_3}
\label{eq10}
\ee
\noindent where $u = u_1 + i u_2$ is a complex-valued function, Eqs.
(\ref{eq8}) and (\ref{eq9}) can be conveniently written as,
\be
(\partial_1 + i A_1 ) u = \mp i (\partial_2 + i A_2 ) u, \\\\\\\
F_{12} = \pm \frac{8 {\mid u \mid}^4}{(1 + {\mid u \mid}^2)^3} .
\label{eq11}
\ee
\noindent The decoupled equation in terms of $u$ is obtained
away from the zeroes of $u$ as,
\be
\bigtriangledown^{2} ln {\mid u \mid}^2 = \frac{8 {\mid u \mid}^4}{(1 +
{\mid u \mid}^2)^3} .
\label{eq12}
\ee
\noindent No exact solution is known for the Eq. (\ref{eq12}).

In order to study the numerical solutions of the Bogomol'nyi equations,
we choose a rotationally symmetric ansatz for the field variables. Our
choice is \cite{dur}
\bea
& & \phi_1 ({\vec{\rho}}, \theta) = sin f(r) \ cos N \theta, \ \
\phi_2 ({\vec{\rho}}, \theta) = sinf(r) \ sin N \theta,\nonumber \\
& & \phi_3 ({\vec{\rho}}, \theta) = cos f(r),
\ \ \ {\vec A}({\vec{\rho}}, \theta)
= - {{\hat e}_{\theta}} \frac{ N a(r)}{\kappa r} \ .
\label{eq13}
\eea
\noindent where $f(r)$ is an arbitrary function and dimensionless length $r =
\frac{\rho}{\kappa}$. $N$ is an integer and also defines the degree of a
topological soliton as will be seen below.
The Eqs. (\ref{eq8}) and
(\ref{eq9}) after substitution of
(\ref{eq13}) reduce to
\be
{f^\prime}(r) = \pm 2 N \ \frac{a+1}{r} \ sin \frac{f}{2} \ \
cos \frac{f}{2} \ , \ \ {a^\prime}(r) = \pm \frac{2 r}{N}
sin^2 f \ \ sin^2 \frac{f}{2} \ .
\label{eq14}
\ee
\noindent The equations in (\ref{eq14}) with upper sign is related
to those with
lower sign by the transformations $f(r) \rightarrow - f(r)$,
$r \rightarrow r$, $a \rightarrow a$ and $N \rightarrow -N$. Here
we consider the lower sign with positive $N$.

The Eq. (\ref{eq14}) is invariant under the transformation $ f(r) \rightarrow
f(r) + 2 \pi$. So it is enough to study the above equations with $f(r)$ having
any value between $0$ to $2 \pi$. Introducing two new variables
 ${\chi_1}(r) = \pi + f(r)$ and
$ {\chi_2}(r) = \pi - f(r)$ and keeping $a(r)$ unchanged, one can easily check
that the ${\chi_1}(r)$, ${\chi_2}(r)$ and $a(r)$ satisfy the same Eq.
(\ref{eq14}). The implication of this is that
for a particular profile of $a(r)$, the solutions
for $f(r)$ are symmetric about $f(r)= \pi$.
Thus we can further restrict the asymptotic values of $f(r)$ between $0$ and
$\pi$. Once a solution is presented within this interval, it automatically
follows that there is a symmetric solution around $f(r)= \pi$ in the interval
$\pi$ to $2 \pi$. Also note that the right hand side of Eq.
(\ref{eq14}) for the gauge field is always negative
( in the case of lower sign ).
So, $a(r)$ is a
decreasing function independent of what specific boundary condition we
choose for the field variables.

The regularity of the field variables near the origin demands that
for finite energy solutions $f(0)= \pi$ and $a(0)=0$. However at the
infinity $f(r)$ can take the value either $0$ or $\pi$ with $a(r)$
approaching some constant. The topological charge for the former
case is $N$, an integer. As a result, the stability of the solutions
for these boundary conditions is of topological nature. However,
when $f(r)$ approaches $\pi$ at infinity, the topological charge defined
in (\ref{eq7}) is not an integer. So, no topological arguments can be used to
establish the stability for solutions with these boundary conditions.
At this point note that the two different conditions
on $f(r)$ at infinity are possible only because of the particular form of the
potential. It is worth pointing out that a similar situation also occurs in
self-dual pure CS theory
\cite{hong}.
In fact, as we shall see below, the profiles of $f(r)$ and $a(r)$ for
$f \rightarrow \pi$ at infinity
is similar to the profiles of the corresponding field variables for the
nontopological solitons in the self-dual pure CS theory. Hence, we refer
these solutions ( $ f(r) \rightarrow \pi$ at infinity )
as nontopological solitons.

Let us first study the profiles of field variables for topological solitons.
The boundary conditions are
\be
f(0) = \pi, \ \ a(0) = 0, \ \ f(r \rightarrow \infty) = 0, \ \
a(r \rightarrow \infty) = - \beta_1
\label{eqbc}
\ee
\noindent Near the origin, i.e. near $f=\pi$,
Eq. (\ref{eq14}) in terms of ${\chi_2}(r) = \pi - f(r)$ reduces
to Liouville equation\footnote{All the solutions, near the origin and at
infinity, given in
this paper are obtained by neglecting terms of the order of ${\chi_2}^3$
in Eq. \ref{eq14}.}.
Hence for small $r$, ${\chi_2}(r)$ can be approximated as,
\be
{\chi_2}(r) = \sqrt{2} (N+1) \left ( \frac{r_0}{r} \right )
\left [ \left (\frac{r_0}{r} \right)^{N+1}
+ \left ( \frac{r}{r_0} \right )^{N+1} \right ]^{-1}
\label{eq15}
\ee
\noindent with the leading behaviour being ${\chi_2}(r) = a_0 r^N$ where
$a_0$ is related to the constant $r_0$ in (\ref{eq15}). Consequently,
the gauge field
$a(r)$ behaves near the origin as,
\be
a(r) = - \frac{2(N+1)}{N} \left ( \frac{r}{r_0} \right )^{N+1}
\left [ \left (\frac{r_0}{r} \right)^{N+1}
+ \left ( \frac{r}{r_0} \right )^{N+1} \right ]^{-1}
\label{eq16}
\ee
\noindent with the leading behaviour being $a(r)= b_0 r^{2(N+1)}$ where
$b_0$ is again related to $r_0$. At infinity, the behaviour of
${\chi_2}(r)$ and $a(r)$ are,
\bea
& & \chi_2 ( r ) = \pi + c_0 r^{- N ( 1- \beta_1 )} +
c_1 r^{-5 N ( 1- \beta_1 ) + 2}
+ O \left ( r^{- 9 N ( 1 - \beta_1 ) + 4 } \right )\nonumber \\
& & a ( r ) = - \beta_1 + d_0  r^{- 4 N ( 1- \beta_1 ) + 2} +
d_1 r^{-8 N ( 1- \beta_1 ) + 4}
+ O \left ( r^{- 12 N ( 1 - \beta_1 ) + 6 } \right )
\label{eq17}
\eea
\noindent where $c_0$, $c_1$, $d_0$ and $d_1$  are arbitrary constants.
It follows from the above equations that $\beta_1 < 1 -
\frac{1}{2 N}$ in order to have nonsingular field variables. As a consequence
, $- 1 < a(r) \leq 0$ for solitons of any degree, since $a(r)$ is a decreasing
function of $r$ and $a(0)= 0$. Since $ -1 < a(r) \leq 0$, it
immediately follows from the first equation of (\ref{eq14}) that $f(r)$
is a decreasing function of $r$. We
have integrated Eq. (\ref{eq14}) numerically for $N=1$ and $2$. Solutions
for $f(r)$ and $a(r)$ indeed exist for any $\beta_1$ in the interval
$0 < \beta_1 < 1 -\frac{1}{2 N}$. The details will be published elsewhere
\cite{gg}.

The topological solitons are characterized by the energy $E= 4 \pi N$,
magnetic flux $\Phi= 2 \pi N \beta_1$, charge $ Q = - \kappa \Phi$
and angular momentum $j_z= \pi \kappa N^2 \beta_1 (2- \beta_1 )$. Note
that though the energy is quantized, the magnetic flux, charge and angular
momentum are not. Thus, for a fixed $N$ there are a family of solutions
characterized by the parameter $\beta_1$ which can take any value
between $0$ and $1 - \frac{1}{2 N}$. This essentially implies that these
solutions are infinitely degenerate. This is reminiscent of degenerate
topological vortex solutions in a generalized Maxwell-CS theory considered
in Ref. \cite{piju}.

The boundary conditions for the nontopological solitons are,
\be
f(0)= \pi, \ \ \ a(0)= 0, \ \ \ f(r \rightarrow \infty)= \pi,
\ \ \ a(r \rightarrow \infty)= -\beta_2.
\label{nt}
\ee
The behaviour of $f(r)$ and $a(r)$ near origin for the nontopological
solitons are still
given by (\ref{eq15}) and (\ref{eq16}) respectively.
The behaviour of the field
variables at infinity can be approximated as,
\bea
& & \chi_2 ( r ) = \sqrt{2} \delta \left ( \frac{r_1}{r} \right )
\left [ \left (\frac{r_1}{r} \right)^\delta
+ \left ( \frac{r}{r_1} \right )^\delta \right ]^{-1}\nonumber \\
& & a(r) = - \beta_2 + \frac{2}{N} \left ( \frac{r_1}{r} \right )^\delta
\left [ \left (\frac{r_1}{r} \right)^\delta
+ \left ( \frac{r}{r_1} \right )^\delta \right ]^{-1}
\label{18}
\eea
\noindent where $\delta=N \beta_2 - N -
1$. The behaviour of $a(r)$ at infinity demands that $\beta_2 > 1 +
\frac{1}{N}$. However, stronger lower bound on $\beta_2$ can be put by
using the following arguments. ${\chi_2}(r)$ solves Liouville equation
with the solution as given in (\ref{eq15}) in the limit ${\chi_2}(r) << 1$.
The corresponding $a(r)$ takes the value $-2 (1+\frac{1}{N})$ at infinity.
So the lower bound on $\beta_2$ follows immediately \cite{jlw}
as $\beta_2 \geq 2 (1+\frac{1}{N})$. It is worth mentioning
at this point that similar arguments can not be valid for topological
solitons. The reason being that the condition ${\chi_2}(r) << 1$ all
over the space does not hold for any soliton solutions in that case.
Because of the lower bound on $\beta_2$, $a(r) + 1$ is no more positive
definite and hence
as we go away from the origin, $f(r)$ decreases up to some point
$r=R$ where $a(R) = -1$ and then increases for $r > R$ reaching
$\pi$ at infinity.

We have integrated Eq. (\ref{eq14}) numerically with the boundary
conditions (\ref{nt}) given above for nontopological solitons. The profile of
$f(r)$ is plotted in Fig. $1$ for $N= 1$ with $\beta_2=4.23$, $5.41$
and $12.25$. The magnetic
field $B(r)= - F_{12}$ is plotted in Fig. 2 for the same values of
$\beta_2$ and $N$.
Notice that for $\beta_2= 5.41$ and $12.25$, the magnetic field has a doubly
degenerate maxima, while for $\beta_2=4.23$ there is no such degeneracy.
This can be explained as follows. The magnetic field written in terms
of $f(r)$ using the second equation of (\ref{eq14}) reads as,
\be
B = \left ( 1 + cos f(r) \right ) \left ( 1 - cos f(r) \right )^2
\label{eqm}
\ee
\noindent One can easily check that at the point of minimum of $f(r)$, say
at $r = \tilde{r}$, the magnetic field becomes maximum if
$cos f(r) > - \frac{1}{3}$. On the other hand, $r=\tilde{r}$ is a local
minimum of $B(r)$ if $cos f(r)$ at that point is less than $-\frac{1}{3}$.
Also the point corresponding to $cos f(r) = - \frac{1}{3}$ is a point of
maximum for the magnetic field. Now notice from Fig. 1 that for
$\beta_2=4.23$, $cos f(r)$ is greater than $-\frac{1}{3}$
( i.e. $f > 1.91$ )
all over space. So, in this case the minimum of $f(r)$ corresponds to the
maximum of $B(r)$. However, for $\beta_2= 5.41$ and $12.25$ the minimum
of $f(r)$ occurs below $1.91$. As a result, the minimum of $f(r)$
corresponds to local minimum of $B(r)$. The maximum of $B(r)$ occurs at that
point for which $cos f(r) = - \frac{1}{3}$. Now observe that this is true for
two different values of $r$. Hence, the maxima of $B(r)$ is doubly
degenerate with a local minimum corresponding to the point of minimum of
$f(r)$. It is also obvious that the maxima of $B(r)$ can be at most doubly
degenerate as $ 0 < f(r) \leq \pi$ for any solution with arbitrary degree $N$.
We have checked that the distance between these two points of maxima
increases as we take higher and higher values of $\beta_2$. We found
numerically that the energy density also has two maxima which are however
not degenerate with the absolute maxima occurring at the point
nearer to the origin.
The electric field is also quite different from that of the
pure CS case \cite{gg}.

The nontopological solitons are characterized by the energy $E= 4 \pi
N \beta_2$, magnetic flux $\Phi= 2 \pi N \beta_2$, charge $Q=
-\kappa \Phi$ and angular momentum $j_z=  \kappa N^2 \beta_2 ( 2- \beta_2 )$.
For these solutions, the energy per unit
charge is given by ${E \over {\mid Q \mid}} = m$, where
$m=\frac{2}{\kappa}$ is the mass of the elementary excitation in the theory.
Thus, these nontopological solitons are at the threshold of the stability
against decay into the elementary excitations.

We now show that both topological as well as nontopological soliton
solutions can also be obtained in a gauged sigma model
with both the Maxwell and the CS term. However, a neutral scalar field
interacting with the $O(3)$ field variables in a specific manner is
necessary in order to have Bogomol'nyi bound. The model we consider
is given by,
\bea
& & {\cal{L}}_1 = \frac{1}{2} D_\mu {\vec \phi}. D^\mu {\vec \phi}
+ \frac{\lambda^2}{2} \partial_\mu \psi \partial^\mu \psi
-\frac{\lambda^2}{4} F_{\mu \nu} F^{\mu \nu} +
\frac{\kappa}{4} \epsilon^{\mu \nu \rho} A_\mu F_{\nu \rho}\nonumber \\
& & - \psi^2  \left ( 1+{\hat{n}}_3 . {\vec \phi} \right ) \left ( 1 -
{\hat{n}}_3 . {\vec \phi} \right )
- \frac{1}{2 \lambda^2} \left ( 1 - \kappa \psi -
{\hat{n}}_3 . {\vec \phi} \right )^2
\label{eq19}
\eea
\noindent where $\psi$ is a neutral scalar field and $\lambda$ is a constant
with dimension of inverse mass. The Bogomol'nyi equations for
static soliton solutions can be obtained following Ref. \cite{lee} as
\bea
& & D_i {\vec \phi} \pm \epsilon_{ij} {\vec \phi} \times D_j {\vec \phi}
= 0,\ \ \ F_{12} \mp \frac{1}{\lambda^2}
\left ( 1 - \phi_3 - \kappa \psi \right ) = 0,\nonumber \\
& & A_0 \pm \psi = 0, \ \ \ \partial_i A_0 \pm \partial_i \psi=0  .
\label{eq20}
\eea
\noindent The topological solitons are obtained when $\phi_3 \rightarrow 1$,
$\psi \rightarrow 0$ at infinity, while for nontopological solitons
$\phi_3 \rightarrow -1$ and $\psi \rightarrow {\frac{2}{\kappa}}$
at infinity. The profiles of
the field variables as well as relevant details will be published elsewhere
\cite{gg}.

\acknowledgements

We thank Professor Avinash Khare for valuable discussions and
critically going through the manuscript.

\begin{figure}
\caption{ A plot of $f(r)$ as a function of $r$ for $N=1$
nontopological soliton solutions with
(a) $\beta_2 = 4.23$; (b) $\beta_2 =5.41$ and (c) $\beta_2 = 12.25$.}

\caption{ A plot of the magnetic field $B(r)$ as a function of
$r$ for $N=1$ nontopological
soliton solutions with
(a) $\beta_2 = 4.23$; (b) $\beta_2 =5.41$ and (c) $\beta_2 = 12.25$.}

\end{figure}

\end{document}